\documentclass[twocolumn,nofootinbib,floatfix,aps,prb,citeautoscript,longbibliography]{revtex4-2}
\usepackage[utf8]{inputenc}
\usepackage[T1]{fontenc}
\usepackage{lineno}

\usepackage{graphicx}
\usepackage{color}
\usepackage{array}
\usepackage{dcolumn}
\usepackage{bm}
\usepackage{multirow}
\usepackage{chemformula}

\usepackage{booktabs}
\usepackage{tabularx}
\usepackage{comment}
\usepackage{tikz}
\usepackage{xstring}

\usepackage{graphicx}
\usepackage{amsmath}
\usepackage{float}
\usepackage{orcidlink}
\usepackage[capitalise]{cleveref}

\begin{document}
\newcommand{\bochum}{Research Center Future Energy Materials and Systems of the University Alliance Ruhr and Interdisciplinary Centre for Advanced Materials Simulation, Ruhr University Bochum, Universitätsstraße 150, D-44801 Bochum, Germany}

\newcommand{\antoine}[1]{\textcolor{orange}{#1}}
\newcommand{\dewen}[1]{\textcolor{blue}{#1}}
\newcommand{\haichen}[1]{\textcolor{purple}{#1}}
\newcommand{\toReplace}[1]{\textcolor{purple}{#1}}
\newcommand{\change}[1]{{#1}}
\newcommand{\revision}[1]{#1}

\title{Universal Machine Learning Interatomic Potentials are Ready for Phonons}

\author{Antoine Loew\,\orcidlink{0009-0008-5018-4895}}
\author{Dewen Sun\,\orcidlink{0009-0003-3487-5593}}
\author{Hai-Chen Wang\,\orcidlink{0000-0002-2892-5879}}
\author{Silvana Botti\,\orcidlink{0000-0002-4920-2370}}
\author{Miguel A. L. Marques\,\orcidlink{0000-0003-0170-8222}} 
\email{miguel.marques@rub.de}
\affiliation{\bochum} 

\date{\today}

\begin{abstract}
There has been an ongoing race for the past several years to develop the best universal machine learning interatomic potential. This progress has led to increasingly accurate models for predicting energy, forces, and stresses, combining innovative architectures with big data. Here, we benchmark these  models on their ability to predict \revision{harmonic} phonon properties, which are critical for understanding the vibrational and thermal behavior of materials. Using around 10\,000 ab initio phonon calculations, we evaluate model performance across various phonon-related parameters to test the universal applicability of these models. The results reveal that some models achieve high accuracy in predicting \revision{harmonic} phonon properties. However, others still exhibit substantial inaccuracies, even if they excel in the prediction of the energy and the forces for materials close to dynamical equilibrium. These findings highlight the importance of considering phonon-related properties in the development of universal machine learning interatomic potentials.
\end{abstract}

\maketitle

\section{Introduction}

One of the most impactful applications of artificial intelligence methods to the field of materials science has been the introduction of machine learning interatomic potentials (MLIPs)~\cite{behler_perspective_2016,graser_machine_2018,schmidt_recent_2019,Unke2021}. These are by now capable of delivering energies and forces at the level of density functional theory (DFT), or beyond, at a computational cost that is often several orders of magnitude lower. As such, they are now accelerating or even replacing the expensive DFT calculations, truly enabling the in-silico design and development of complex materials.

Various representation methods for crystal structures (embedding techniques) have been proposed in the past years~\cite{Behler2007,Bartk2013,Gasteiger_graph,batzner_e3-equivariant_2022,Gastegger2021}. These methods encode crystal structure information into learnable features, thereby improving data efficiency for the models. Furthermore, new machine learning models and strategies were also developed and improved. These advancements gained significant momentum with the introduction of message passing neural network frameworks~\cite{Gimler}, which were later enhanced by incorporating continuous-filter convolutions for message passing~\cite{schnet2017}. Message passing addressed the issue of exponentially expanding descriptor sizes in earlier machine learning models, enabling the prediction of much larger and more complex systems. 

The training of MLIPs has also been facilitated by the continuous accumulation of DFT calculations over the decades, and by the creation of comprehensive databases, such as the Materials Project~\cite{Jain2013}, the Open Quantum Materials Database~\cite{Kirklin2015}, Aflowlib~\cite{Curtarolo2012},  Alexandria~\cite{Schmidt2024_1}, NOMAD~\cite{Scheidgen2023}, etc. These databases contain materials with almost all chemical elements and in all types of crystal structures. They also provide a variety of computed properties, including total energies, forces, stresses, etc. not only for compounds at dynamical equilibrium, but also for geometry optimization or molecular dynamics paths.

Until recently, MLIPs were typically trained for specific chemical systems and were often limited to a narrow range of geometries and atomic arrangements. This paradigm shifted in 2019 with the introduction of the Materials Graph Network (MEGNet)~\cite{Chen2019}, a framework designed for \textit{universal} machine learning in materials science. Universal MLIPs (uMLIPs) are foundational models capable of handling all chemistries and crystal structures. MEGNet already demonstrated relatively low prediction errors across a wide array of properties in both molecules and crystals. Its performance was significantly enhanced by incorporating atomic coordinates, lattice vectors in crystals, and 3-body interactions~\cite{chen_universal_2022}, enabling uMLIPs to predict ground-state geometries with a mean absolute error of 0.035~eV/atom in the energy, when compared to DFT. Further advancements, such as the use of higher-order body messages, have resulted in models that are accurate, fast, and highly parallelizable~\cite{batatia_mace_2023}. Since then, there has been a surge of developments, with new and improved models being published at an almost monthly rate~\cite{Neumann2024, Park2024, Liao_2023, Deng2023, Choudhary2021}.

In spite of the rapid progress of uMLIPs, there are still challenges remaining. Since these models are mostly trained and evaluated on existing datasets~\cite{Jain2013, Schmidt2024_1, Deng2023}, containing mainly equilibrium or near-equilibrium geometries, they struggle to reproduce meta-stable or highly distorted structures~\cite{Riebesell_matbench}. To resolve this problem, further information on the off-equilibrium structures from molecular dynamic results can be used~\cite{Liao_nonEq}. Alternatively, by gradually distorting the optimized geometries, one can step away from the minima of the potential energy surface~\cite{Barroso2024}. Models trained on such augmented datasets show superior performance at predicting equilibrium structures and energies~\cite{Barroso2024}. Moreover, compared to those trained without off-equilibrium data, models trained with augmented datasets perform better on predicting the first derivatives of the energy~\cite{Barroso2024}. \change{While multiple evaluations of uMLIPs can be found in the literature~\cite{Focassio2024, pota2024, Haochen2024}, direct phonon prediction capabilities have not been comprehensively characterized.}

Here we benchmark seven uMLIP models, specifically M3GNet, CHGNet, \change{MACE-MP-0}, \change{SevenNet-0}, \change{MatterSim-v1}, ORB, and  \change{eqV2-M}, for the calculation of phonon properties. These properties are obtained from the second derivatives (i.e., the curvature) of the potential energy surface, and therefore sample a small neighbourhood around the dynamically stable minima. Phonons are extremely important in materials science, as they are fundamental in determining the free energy (and therefore thermodynamic stability), dynamical stability, thermal properties, etc. We note that all seven models are also featured in the Matbench Discovery leaderboard~\cite{Riebesell_matbench,Deng2023, Chen2019,batatia_mace_2023,Park2024,Neumann2024,Barroso2024,mattersim} (ranked 12th, 11th, 10th, 8th, 3rd, 2nd, and 1st, respectively, at the time of writing).

M3GNet~\cite{Chen2019} is one of the pioneering uMLIPs and still remains a key model in the field. It employs three-body interactions and incorporates atomic positions, enabling the calculation of forces through the automatic differentiation of the neural network.
CHGNet~\cite{Deng2023} is another of the earlier models, but it still demonstrates excellent performance while having one of the smallest \revision{architectures} with just over 400 thousand parameters.
\change{MACE-MP-0}~\cite{batatia_mace_2023} utilizes the atomic cluster expansion~\cite{Drautz2019} as a local descriptor, reducing the number of necessary message-passing steps while maintaining efficiency.
\change{SevenNet-0}~\cite{Park2024}, built upon NequIP~\cite{batzner_e3-equivariant_2022}, focuses on parallelizing the message-passing process. This approach preserves NequIP's data efficiency, accuracy, and equivariant character.
\change{MatterSim-v1}~\cite{mattersim} builds upon M3GNet, leveraging active learning and efficient sampling across the chemical space. Its goal is to enhance the accuracy of energy and force predictions over a broader range of scenarios while maintaining a straightforward architecture that is easy to fine-tune.
The ORB model~\cite{Neumann2024} combines the smooth overlap of atomic positions~\cite{Bartk2013} with a graph network simulator~\cite{SanchezGonzalez2020LearningTS}.
Finally, \change{eqV2-M}~\cite{Barroso2024} \change{is using the model developed by Ref.~\onlinecite{Liao_2023} utilizing equivariant transformers to achieve higher-order equivariant representations}. An important detail to note is that the ORB and \change{eqV2-M} models predict forces as a separate output rather than deriving them as energy gradients as the other five models.

\section{Results}
\label{sec:results}

\subsection{Dataset and Its Properties}

\begin{figure}[htb!]
  \centering
  \includegraphics[width=.75\columnwidth]{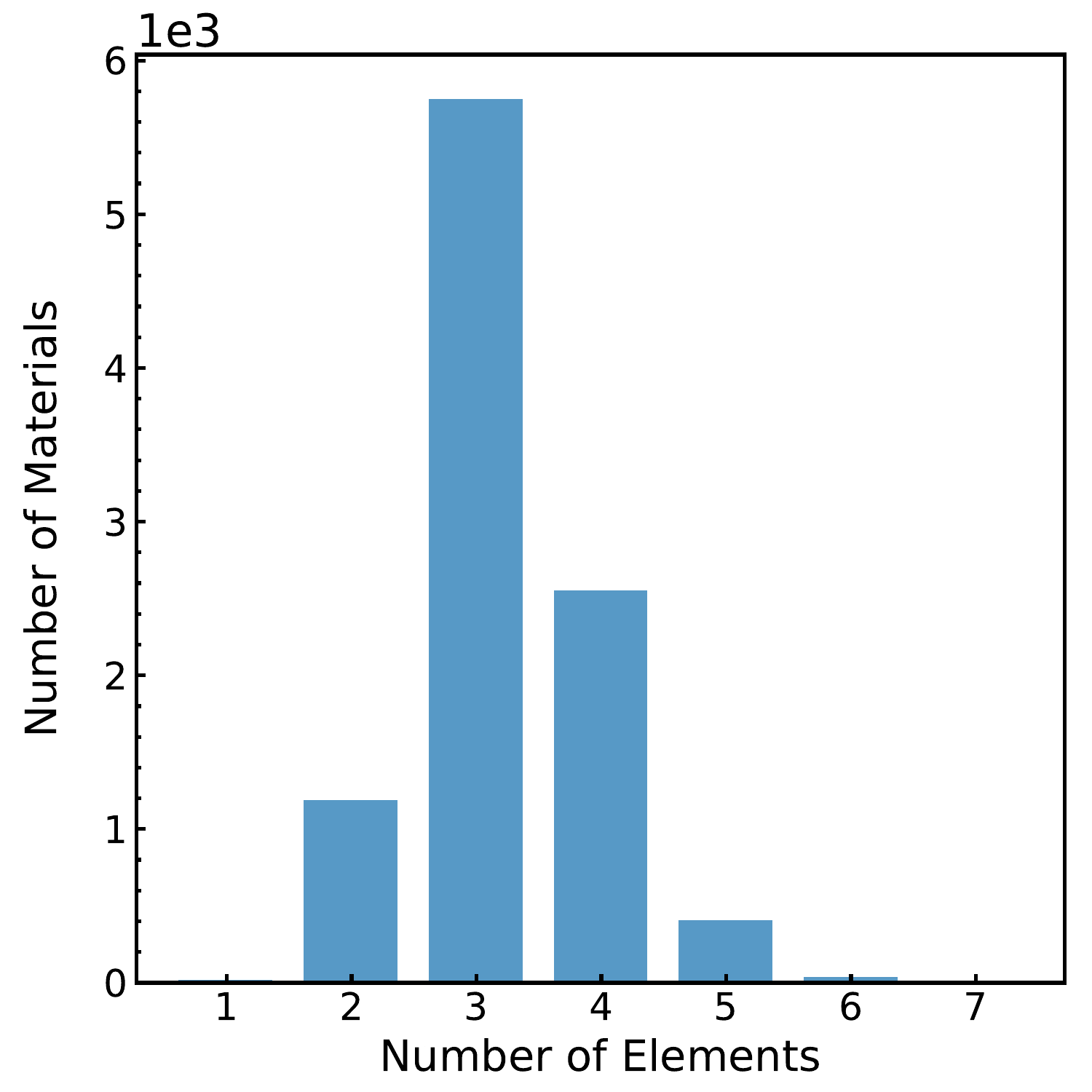}\\
  (a)\\
  \includegraphics[width=.75\columnwidth]{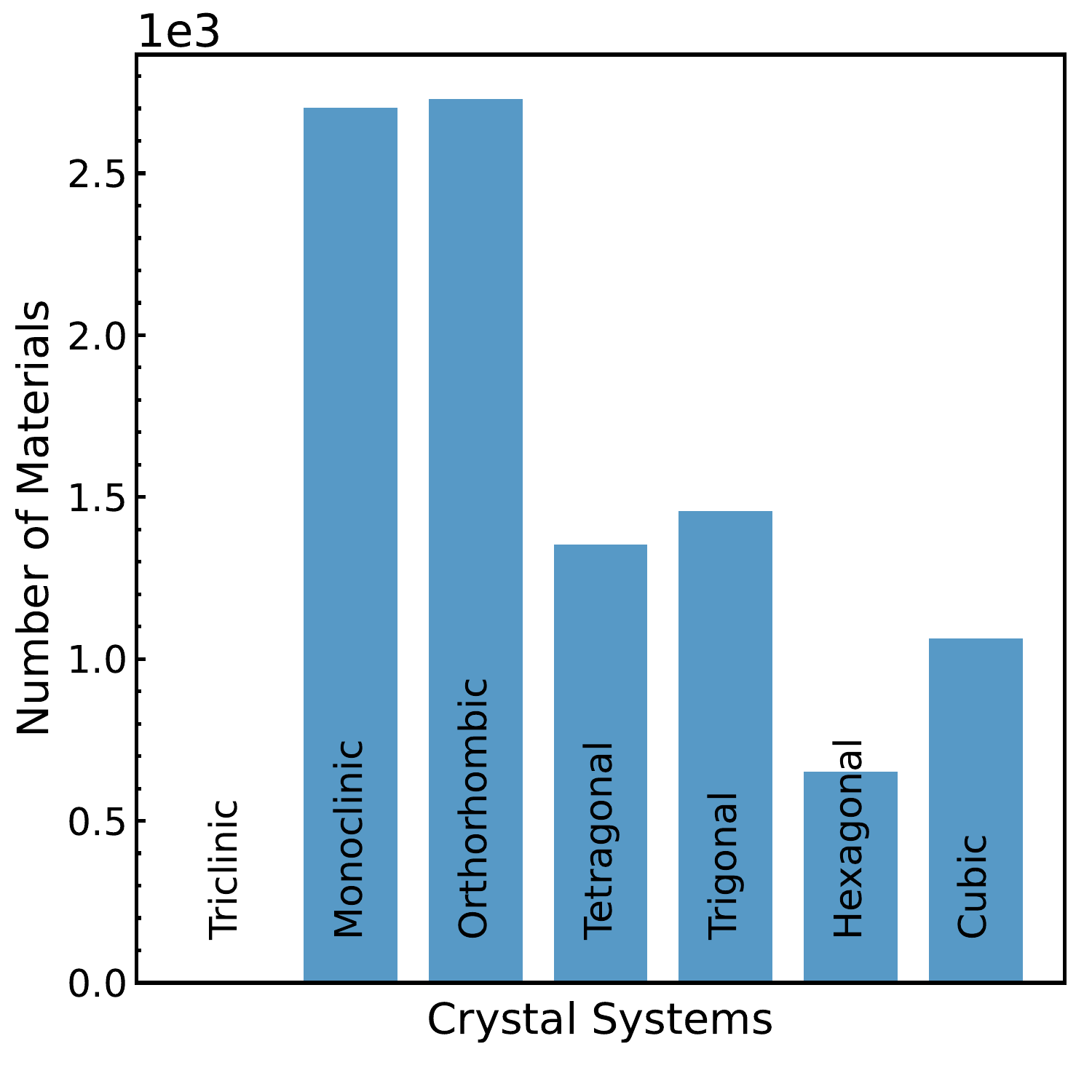} \\
  (b)\\
  \includegraphics[width=0.75\columnwidth]{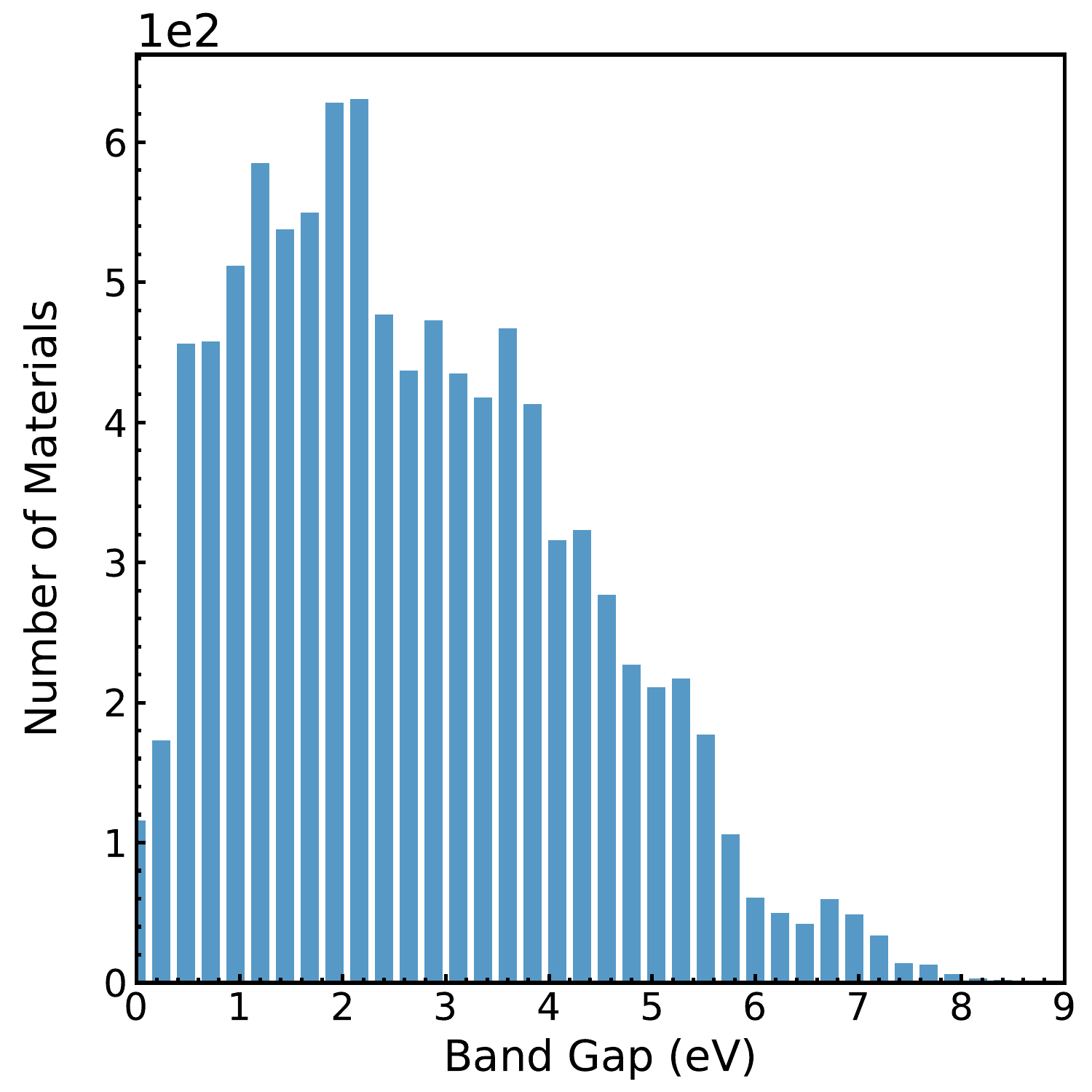}\\
  (c)
  \caption{Distribution of (a)~number of different chemical elements per unit cell, (b)~crystal systems, and (c)~band gaps calculated with the PBE functional for all the materials in the dataset.}
  \label{fig:n_atoms_elements}
\end{figure}

To benchmark phonon properties we use the dataset developed in the MDR database~\cite{MDR_database}. This dataset includes around 10\,000 non-magnetic semiconductors, covering a wide range of elements across the periodic table. Moreover, the phonon calculations were performed with \textsc{vasp}, ensuring a high degree of compatibility with the training sets used in the construction of the uMLIPs. Unfortunately, this phonon dataset was originally constructed with the Perdew-Burke-Ernzerhof (PBE) for solids (PBEsol)~\cite{perdew_restoring_2008} approximation to the exchange-correlation functional. This is certainly a very reasonable choice, as the PBEsol functional exhibits superior structural~\cite{Csonka2009,Hussein_2022} and phonon~\cite{He2014} properties when compared to the standard PBE~\cite{Perdew1996}. However, as all uMLIPs were trained on PBE data, a direct comparison to PBEsol phonons can be ambiguous. To mitigate this problem, we recalculated the entire phonon dataset from Ref.~\onlinecite{MDR_database} with the PBE functional (see \cref{sec:method}). In the following, we not only present comparisons of uMLIP calculations with PBE data, but we also include the difference between PBE and PBEsol. This gives us an estimate of the variability of the results as a function of the approximation to the exchange-correlation function, that we use as an absolute scale to assess the quality of the uMLIPs.

As illustrated in \cref{fig:n_atoms_elements}a, the dataset contains mostly ternary and quaternary compounds. Additionally, we observe that the majority of the compounds belong to the monoclinic and orthorhombic crystal systems, followed by approximately equal proportions of trigonal and tetragonal systems. Cubic systems are less common, with hexagonal systems representing the smallest proportion. Ultimately, these characteristics are inherited from the Materials Project database~\cite{Jain2013} and the Inorganic Crystal Structure Database (ICSD)~\cite{Bergerhoff1983}. Finally, triclinic systems are absent from the MDR database, likely because of the extra computational cost that arises from the reduced symmetry. 

\begin{figure*}[htb]
\centering
\includegraphics[width=0.8\linewidth]{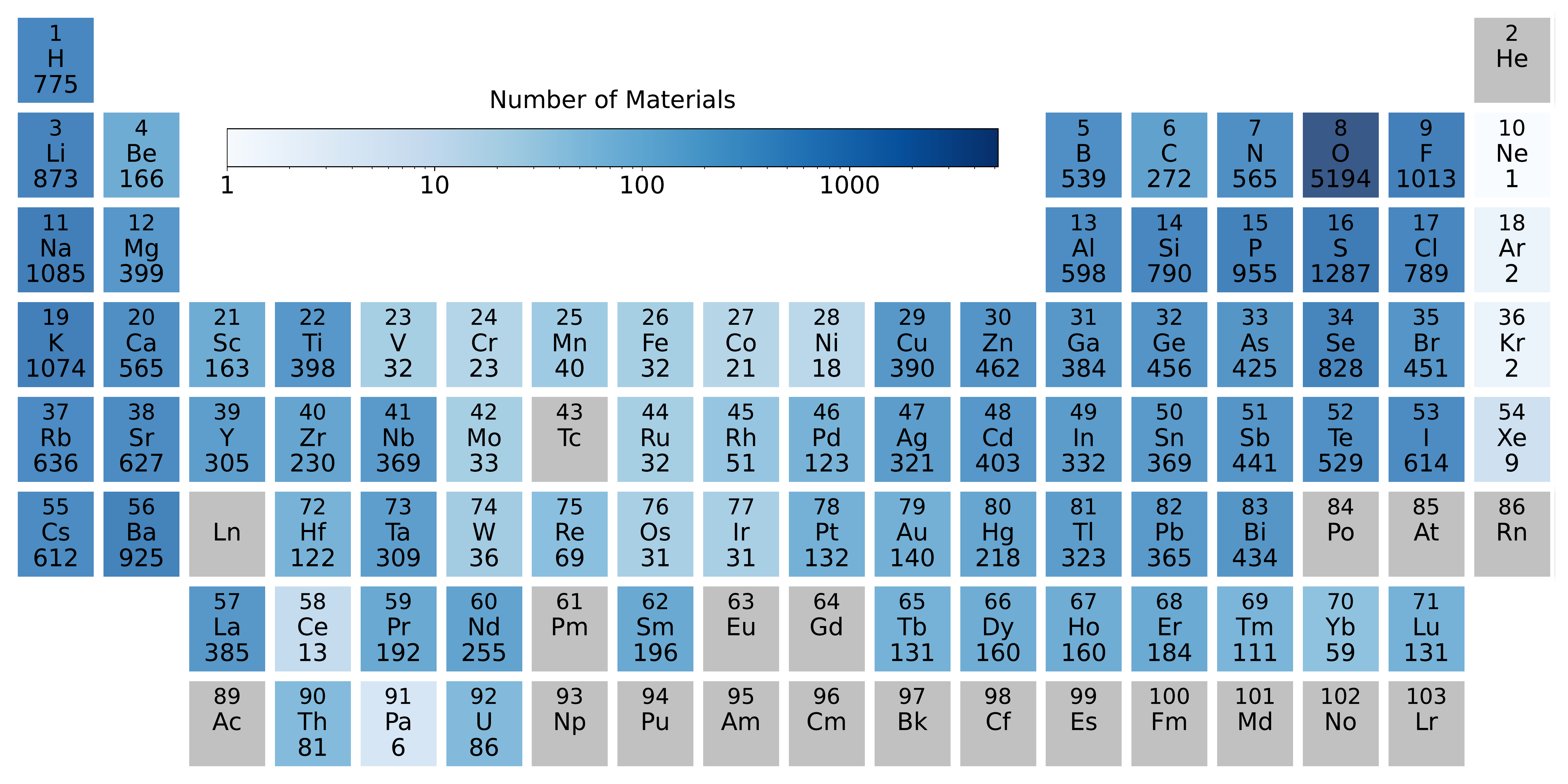}\\
\caption{Periodic tables showing the frequency of the chemical elements in the structures from the dataset. Elements in gray are absent from the dataset.}
 \label{fig:elements}
\end{figure*}

In ~\cref{fig:elements} we plot the frequency of the chemical elements in the dataset. We can see that almost all the periodic table is well represented (with a few exceptions like Tc that is radioactive or Eu and Gd for which \textsc{vasp} has convergence problems). We also observe a significant abundance of structures containing oxygen. However, certain compounds, such as those containing Mo and W, as well as the magnetic 3$d$ elements (from V to Ni) are underrepresented. These biases in the MDR database~\cite{MDR_database} are also to some extent inherited from the Materials Project database~\cite{Jain2013}, but should not be relevant for the benchmark we present here. Although the dataset is predominated by oxides, the band gaps of the whole set still covers a large range, as illustrated in \cref{fig:n_atoms_elements}c.

\subsection{Relative Performances of uMLIPs}

\begin{table}[htp]
\centering
\caption{Summary of the errors for energy ($E$, in~meV/atom) and volume ($V$, in \AA$^3$/atom) of the converged relaxation compared to PBE results. We list mean absolute errors (MAE) as well as the percentage of failed (F) to converge.}
\label{table:summarytraining}
\begin{tabular*}{\columnwidth}{@{\extracolsep{\fill}} l c c c c c}
\toprule
Model& Failed (\%) & MAE($E$)   & MAE($V$)     \\
\midrule
M3GNet    & 0.12   & 33         & 0.516        \\
CHGNet    & 0.09   & 334        & 0.518        \\
\change{MACE-MP-0}      & 0.14   & 31         & 0.392        \\
\change{SevenNet-0}  & 0.15   & 31         & 0.283        \\
\change{MatterSim-v1} & 0.10   & 29         & 0.244        \\
ORB       & 0.82   & 31         & 0.082        \\
\change{eqV2-M}    & 0.85   & 33         & 0.084        \\
PBESol    & --     &  --        & 1.283        \\
\bottomrule
\end{tabular*}
\end{table}

\begin{figure}[htb]
\centering
\includegraphics[width=0.8\columnwidth]{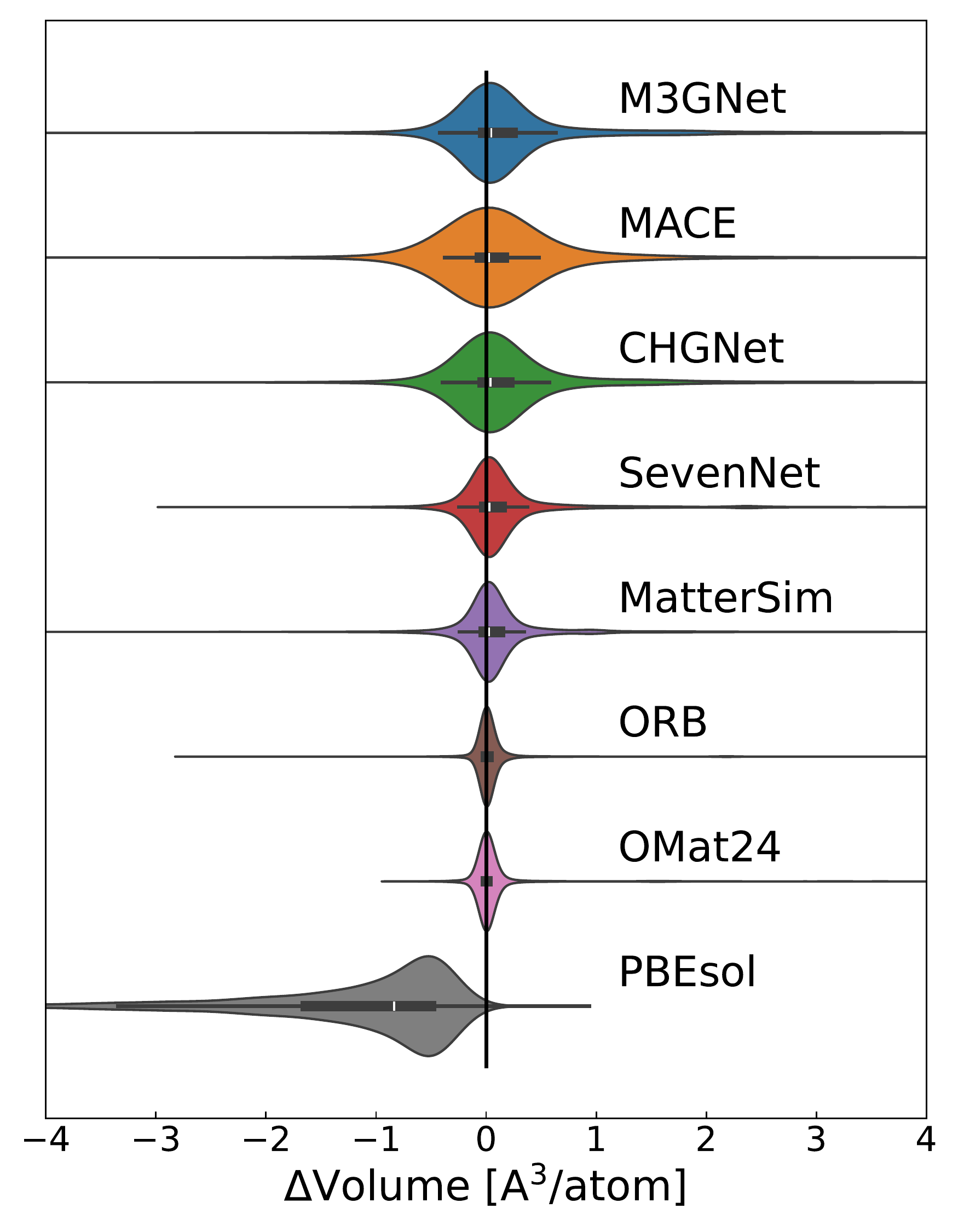}
\caption{Violin plot of the errors in the volume of the unit cell per atoms, relatively to the PBE reference data.}
 \label{fig:voilin_e_v}
\end{figure}

We start by discussing the errors in the geometry relaxations, as shown in \cref{table:summarytraining} and \cref{fig:voilin_e_v}. The ``Failed'' column in \cref{table:summarytraining} indicates for how many systems a model failed to converge the forces to below 0.005~eV/\AA. We can see that CHGNet and \change{MatterSim-v1} models appears to be the most reliable, with respectively 0.09\% and 0.10\% unconverged structures. The M3GNet, \change{SevenNet-0} and \change{MACE-MP-0} models have a similar number of unconverged structures, while the ORB and \change{eqV2-M} models exhibit a much larger failure rate. The most unreliable model for this dataset is \change{eqV2-M}, for which 0.85\% \revision{structural calculations were unable to converge}. In general, there are two main reasons behind the failures, specifically either the geometry optimization path explored regions of the potential energy surface for which the uMLIP yielded unphysical forces, or there were high frequency errors in the forces that prevented the relaxation algorithm to converge to the required precision. This latter reason is behind the very large failure rate for the two models where the forces are not the exact derivatives of the energy. \change{CHGNet shows notably higher error in energy predictions, which is expected given that we did not apply the energy correction procedure typically used during CHGNet's training.}

Looking at \cref{fig:voilin_e_v} we see that, as expected, PBEsol leads to a contraction of the unit cell, correcting the underbinding that is typical of the PBE approximation. The large majority of the systems show a difference between the PBE and PBEsol volume per atom between 0 and -2~\AA$^3$/atom. All uMLIPs exhibit a MAE($V$) that is smaller than the mean absolute difference between PBE and  PBEsol. Among them, the \change{eqV2-M} model emerges as the most accurate, closely followed by ORB. Indeed, these two uMLIPs show a remarkable performance for the vast majority of the compounds in the dataset, with errors that are quite small in both absolute and relative terms. \change{MatterSim-v1} and \change{SevenNet-0} show a very solid performance, although with mean errors four times larger than the two best models. Finally, M3GNet, \change{MACE-MP-0}, and CHGNet have a wider error distribution, with  a MAE in the range of 0.4--0.5~\AA$^3$/atom. \revision{These results confirm} that both \change{eqV2-M} and ORB are the best models for geometry optimization, and that they can already be used to essentially replace DFT calculations for this task.

We now turn our attention to phonon related properties. We chose to look at the maximum phonon frequency \revision{(reported in Kelvin, with $1~\mathrm{K}=0.695$~cm$^{-1}$)}, \revision{the phonon density of state (DOS), the average of the sound velocity on the 3 accoustic branches,} the vibrational entropy,  the Helmholtz free energy, and the heat capacity at constant volume, the last three calculated at the temperature of 300~K. The maximum phonon frequency allows us to detect systematic errors in the prediction of the concavity of the potential energy surface, especially important as it is well known that some uMLIPs have the tendency to yield too soft phonons. \revision{The phonon DOS provides information regarding the general prediction of phonon modes with respect to frequency, while the sound velocity help identify errors in the acoustic branches in the vicinity of $\Gamma$. It should be noted that for the phonon DOS we remove values below 0.1 states/THz.} The vibrational entropy and the Helmholtz free energy are important properties as they are essential to determine thermodynamic stability and phase diagrams as a function of temperature. Finally, the heat capacity is an important thermal property that can be directly measured experimentally.

We note that maximum phonon frequency was calculated from the values at the $q$-points commensurate to the supercell matrix, whereas \revision{the DOS and} thermodynamic properties were obtained on an denser $q$-grid by applying Fourier interpolation (see~\cref{sec:method}). However, as the $q$-grids are consistent across DFT and uMLIPs calculations, the interpolation error should be systematic and should not affect the benchmark.

\begin{table}[htb!]
\centering
\caption{Summary of the mean absolute errors (MAE) for the maximum phonon frequency (MAE($\omega_\text{max}$), in \revision{Kelvin where \(1~\mathrm{K}{\approx}0.695~\mathrm{cm}^{-1}\)}), the vibrational entropy (MAE($S$), in J/K/mol), the Helmholtz free energy (MAE($F$), in kJ/mol), the heat capacity at constant volume (MAE($C_V$), in J/K/mol)\revision{, the phonon density of state (MAE($DOS$)), and the average of sound velocties (MAE($avg.v_\text{s}$))}}
\label{tab:error_phonon}
\begin{tabular*}{\columnwidth}{@{\extracolsep{\fill}} l c c c}
\toprule
Model       & MAE($\omega_\text{max}$)& MAE($S$)  & MAE($F$)\\
\midrule
M3GNet      & 98                  & 150  & 56    \\
CHGNet      & 89                  & 114  & 45   \\
\change{MACE-MP-0}        & 61                  & 60   & 24   \\
\change{SevenNet-0}    & 40                  & 48   & 19  \\
\change{MatterSim-v1}   & 17                  & 15   & 5    \\
ORB         & 291                 & 421  & 175  \\
\change{eqV2-M}      & 780                 & 403  & 241  \\
PBESol      & 33                  & 25   & 10\\
\bottomrule
\end{tabular*}
\vspace{1em}

\begin{tabular*}{\columnwidth}{@{\extracolsep{\fill}} l c c c}

\toprule
\revision{Model}   &   \revision{MAE($C_V$)}  & \revision{MAE($DOS$)} & \revision{MAE($avg.v_\text{s}$)}\\
\midrule
M3GNet              & 22      & 5.5  & 617    \\
CHGNet               & 21   & 4.8  & 649     \\
\change{MACE-MP-0}  & 13   & 4.2   & 523     \\
\change{SevenNet-0}  & 9       & 4.2   & 510    \\
\change{MatterSim-v1}  & 3      & 3.3   & 401   \\
ORB                  & 57    & 9.1  & 1198  \\
\change{eqV2-M}      & 100    & 10.7  & 1240 \\
PBESol               & 5     & 2.9   & 305 \\

\bottomrule

\end{tabular*}

\end{table}

\begin{figure*}[htp!]
\centering
\begin{tabular}{c c c}
\includegraphics[width=0.32\linewidth]{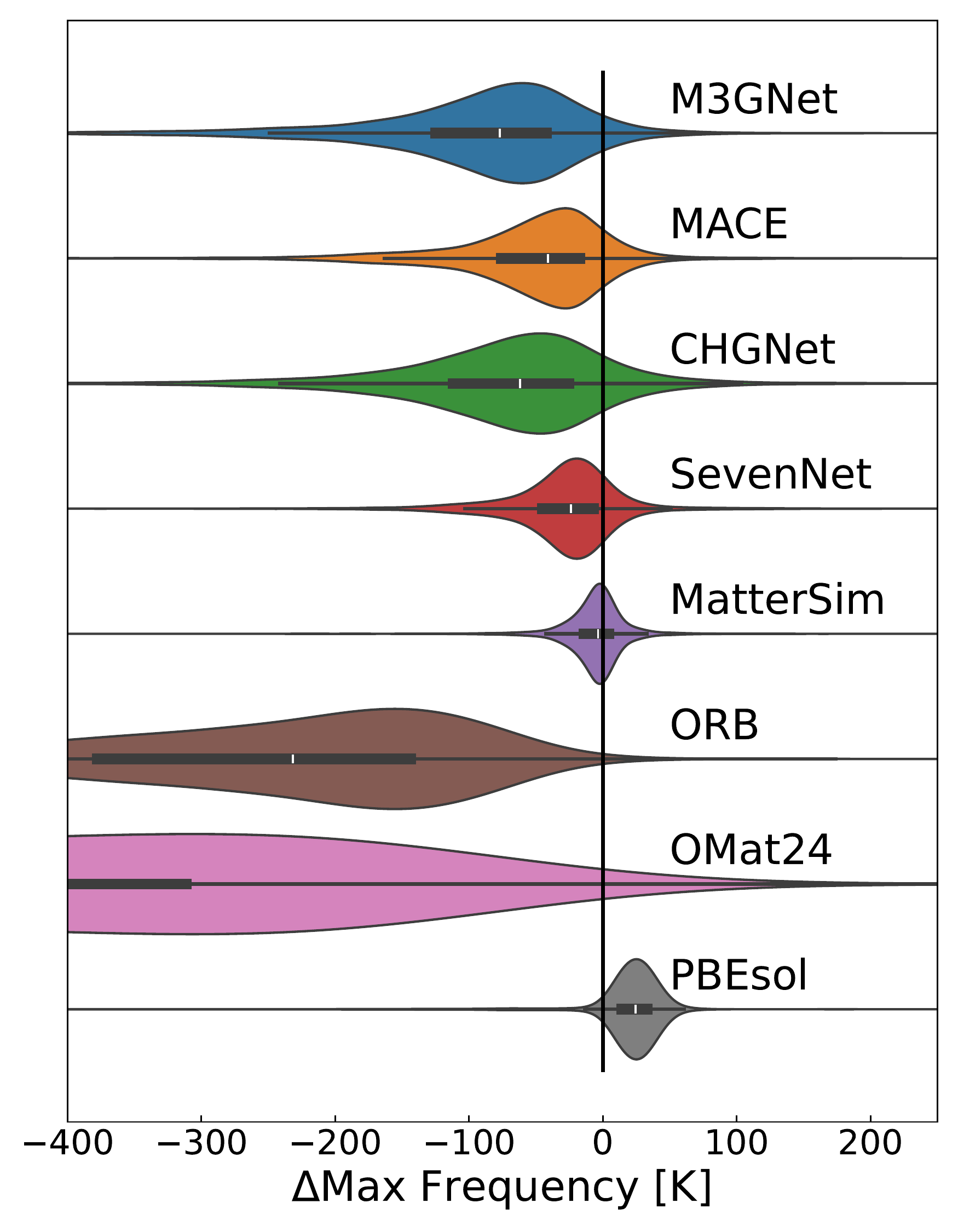} &
\includegraphics[width=0.32\linewidth]{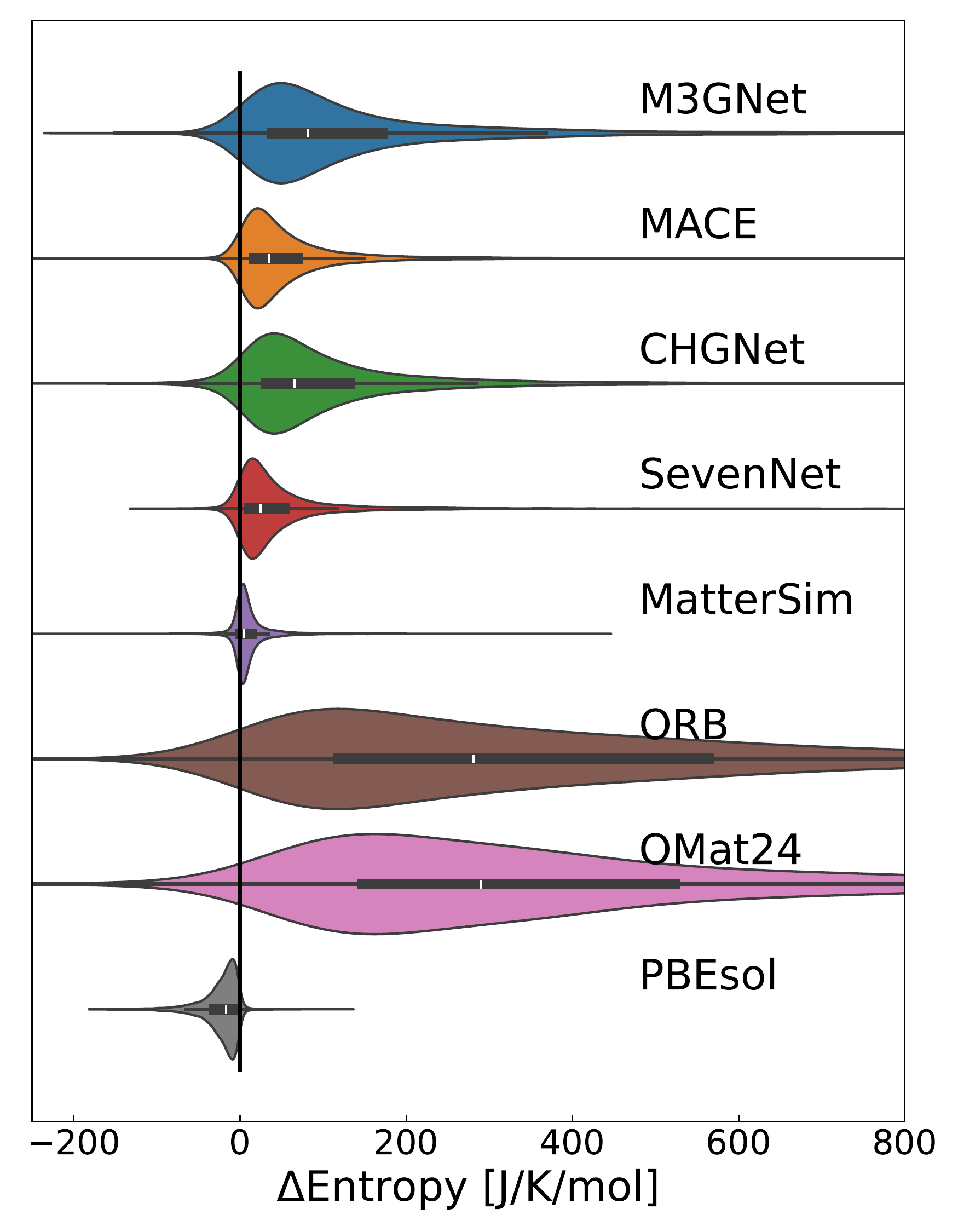} &
\includegraphics[width=0.32\linewidth]{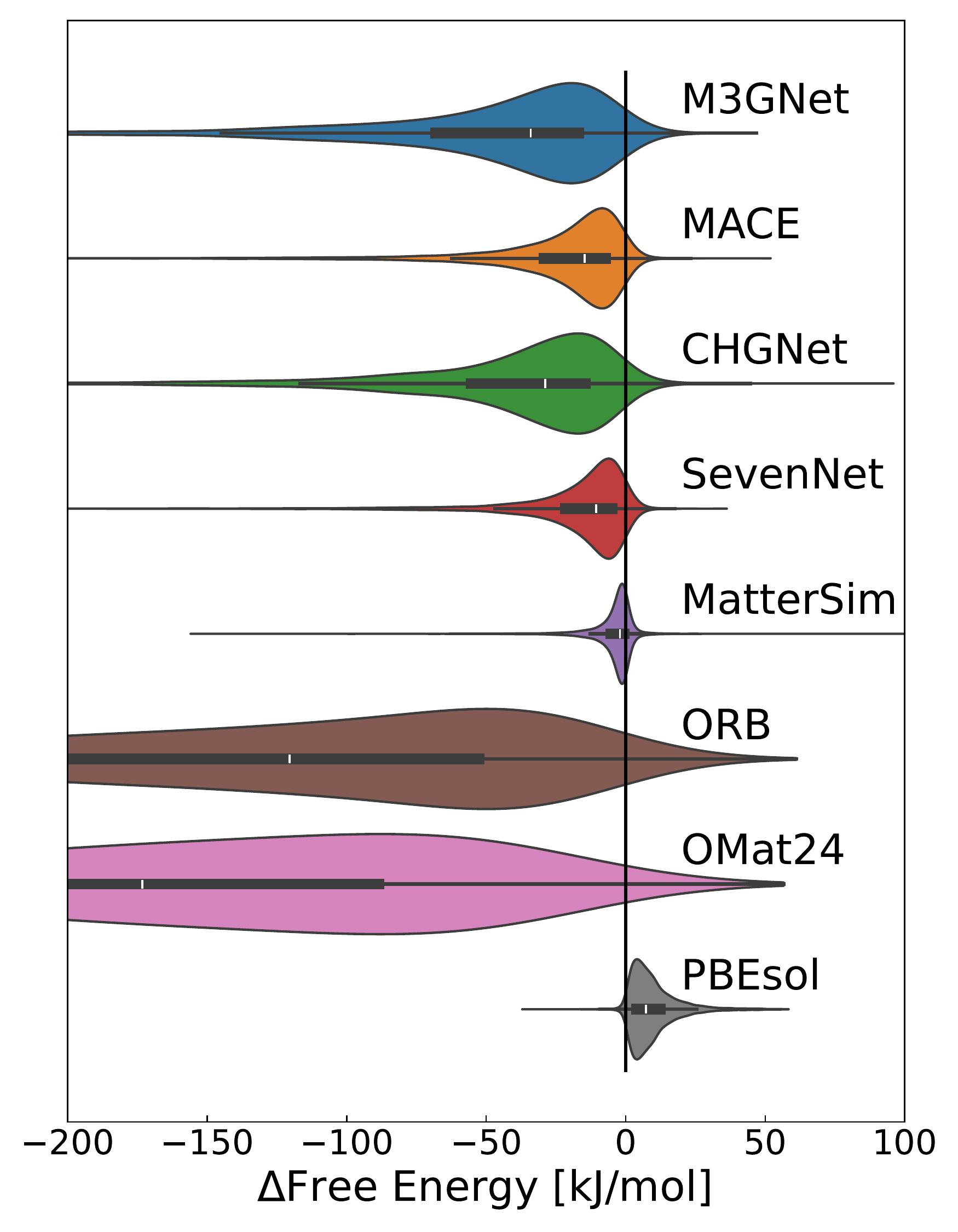} \\
(a) & (b) & (c) \\
\includegraphics[width=0.32\linewidth]{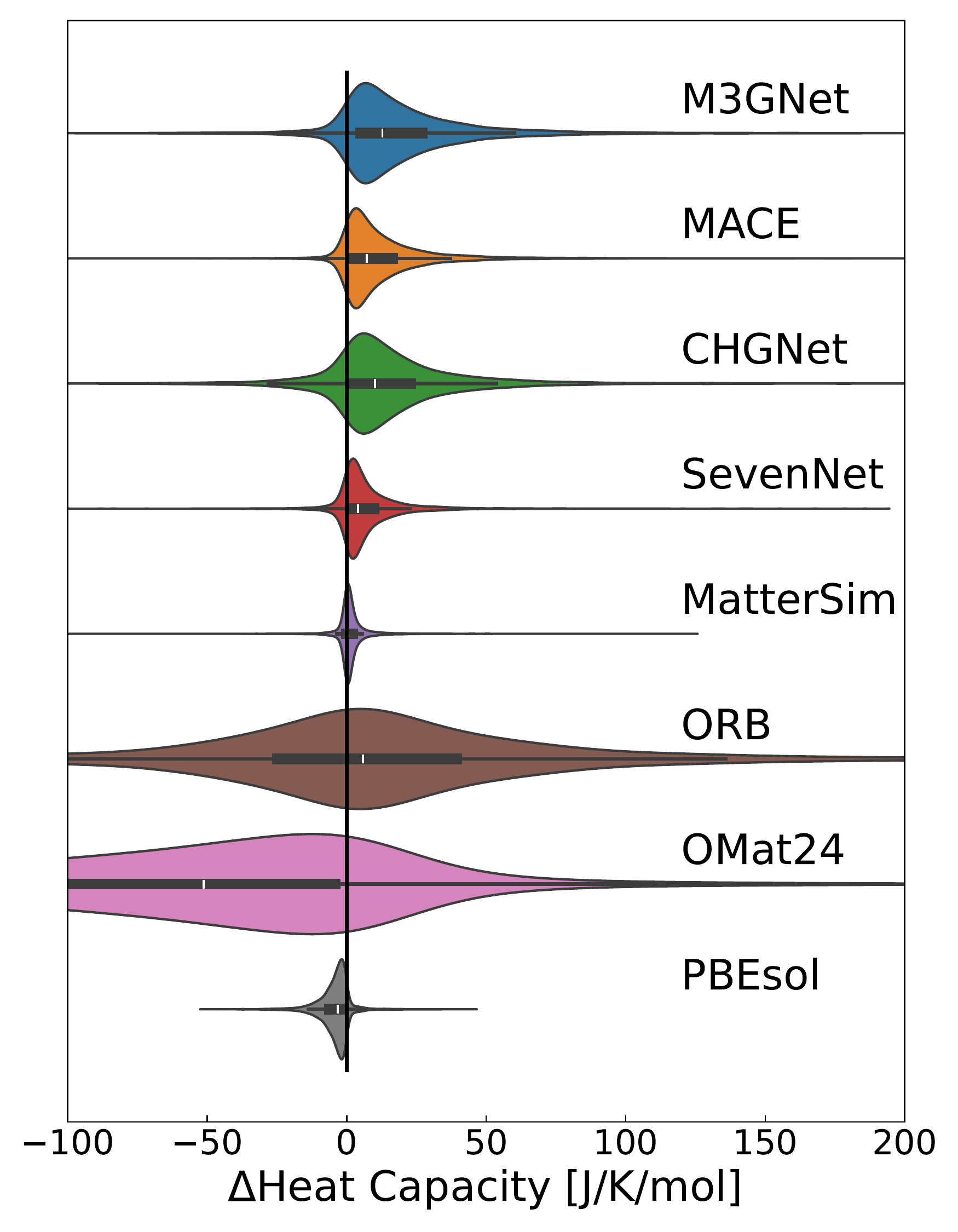} & 
\includegraphics[width=0.32\linewidth]{img/stats/dos_violin_plots_newlabels.pdf} &
\includegraphics[width=0.32\linewidth]{img/stats/avg_sound_velocity_violin_plots_newlabels.pdf}\\
(d) & (e) & (f) \\
\end{tabular}
\caption{Violin plots of the errors in (a)~the maximum phonon frequency, (b)~the vibrational entropy, (c)~the Helmholtz free energy, (d)~the heat capacity, \revision{(e)~the density of states and (f)~the average of the sound velocity on the 3 accoustic branches}, relatively to the PBE reference data.}
 \label{fig:violin_phonon}
\end{figure*}

\begin{figure*}[htb!]
\centering
\includegraphics[width=0.9\linewidth]{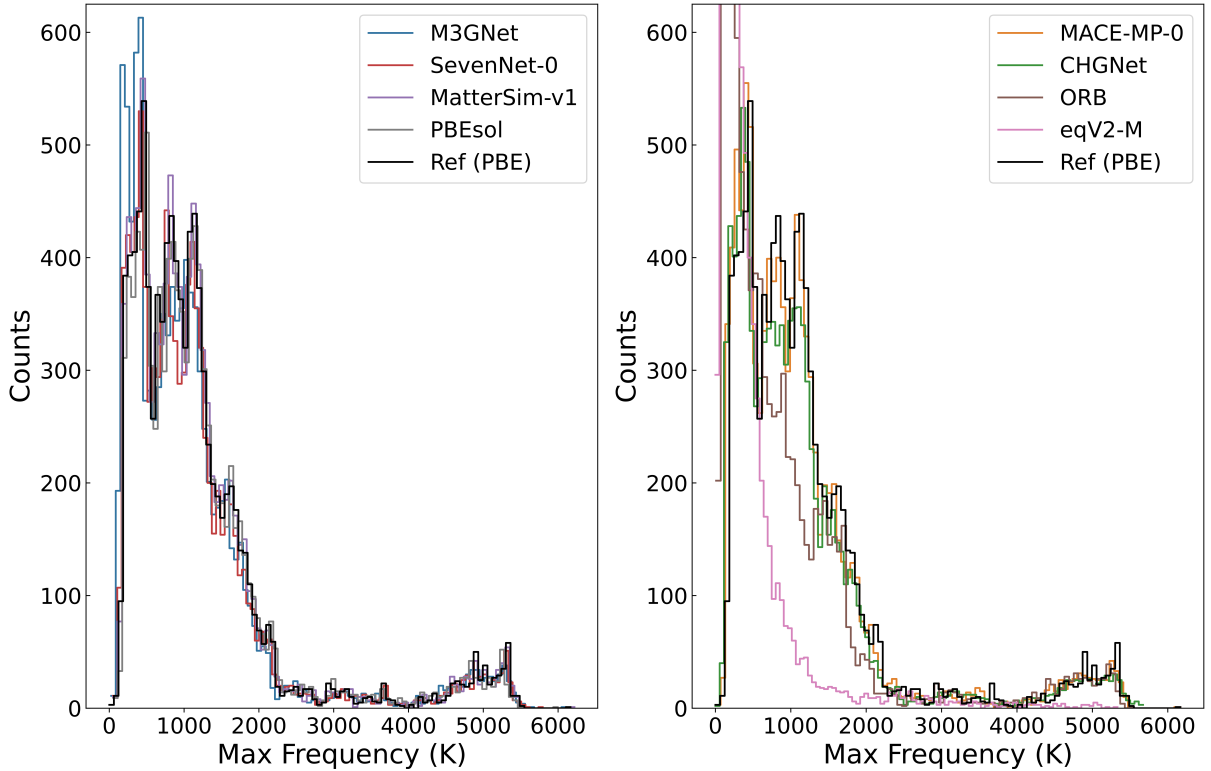}
\caption{Highest frequencies predicted for each structure for all models and from the original PBEsol MDR database.}
 \label{fig:max_freq}
\end{figure*}

We aggregated the errors for all models in \cref{tab:error_phonon} and in \cref{fig:violin_phonon}. We first notice that the deviation between the PBE and PBEsol results is small but not negligible. This observation reinforces the necessity of using a consistent functional between the training and benchmarking stages. The difference between PBE and PBEsol exhibits a rather narrow distribution in all \revision{6} properties, especially when compared to the MAE of most of the uMLIPs. There are also systematic differences: for example, the maximum phonon frequencies in PBEsol are higher than with PBE, which can be understood by the contraction of the cell and subsequent hardening of the force constants. PBEsol also leads to larger values of the free energy (on average of the order of 10~kJ/mol), and to smaller values of the entropy and the heat capacity. 

Based on the errors we can roughly classify the seven models into three categories. The first contains ORB and \change{eqV2-M}, which have very large errors in phonon-related properties (see \cref{fig:violin_phonon}). In fact, phonon frequencies are grossly underestimated, and are often even imaginary as we will see in the following. 

In the second category we have, in increasing order of accuracy, M3GNet, CHGNet, \change{MACE-MP-0} and \change{SevenNet-0} (see \cref{fig:violin_phonon}). The errors of these models are on average considerably larger than the difference between PBE and PBEsol. Moreover they all exhibit systematic errors, underestimating the phonon frequencies and the free energy, and overestimating the entropy and the heat capacity. From the four models, the most accurate is clearly \change{SevenNet-0}, while the older M3GNet and CHGNet show the larger errors. In spite of the difference in topologies, these four models are all trained in the same dataset, so it is not surprising that their results are somewhat similar. This again demonstrates that training data is at least as important as the representation of the crystal structure or the topology of the model to develop a uMLIP.

Finally, \change{MatterSim-v1} stands out as the most accurate uMLIP for the calculation of phonons. \revision{Not only does it} not exhibit any strong systematic error, with all distributions essentially centered at zero, but also the dispersion of the errors is extremely small, leading to values of MAE considerably smaller than the difference between PBE and PBEsol. This indicates that \change{MatterSim-v1} can be used to calculate phonon properties of semiconductors with an accuracy comparable to DFT codes, although at a very small fraction of the computational cost. It is very interesting to note that although \change{MatterSim-v1} is based upon the simple M3GNet, its performance exceeds much more complicated models such as \change{SevenNet-0} or \change{eqV2-M} that are based on equivariant networks. The key in this case is the scalability of M3GNet, which allows for an increase in the number of parameters and the efficient use of larger amounts of training data.

To have a better understanding of the general behavior of the uMLIPs, we plot in \cref{fig:max_freq} the distribution of the maximum frequency predicted. Most compounds have maximum frequencies in the range of 500--2000~K, with a few containing very light elements going up to 5500~K. The softening of the phonon frequencies by M3GNet, CHGNet, \change{MACE-MP-0} and \change{SevenNet-0} is evident, in particular for the first two. ORB and \change{eqV2-M}, on the other hand, exhibit completely distorted distributions peaking at zero, showing that the force constants obtained with these models are unphysical.

\begin{table}[htb!]
    \caption{Normalized confusion matrix with true stable (TS, in \%), false unstable (FU, in \%), true unstable (TU, in \%), false stable (FS, in \%) of predicted dynamical stability for models using PBE data as reference. For information, the number of stable compounds in the PBE is 8189 and unstable is 1769.}
    \label{tab:confusionmatrix}
    \centering
    \begin{tabularx}{\columnwidth}{@{\extracolsep{\fill}} l c c c c }
\toprule
Model       & TS        & FU        & TU        & FS    \\
\midrule
M3GNet      & 87        & 13        & 73        & 27    \\
CHGNet      & 77        & 23        & 73        & 27    \\
\change{MACE-MP-0}        & 95        & 5         & 73        & 27    \\
\change{SevenNet-0}    & 81        & 19        & 80        & 20    \\
\change{MatterSim-v1}   & 95        & 5         & 75        & 25    \\
ORB         & 15        & 85        & 92        & 8     \\
\change{eqV2-M}      & 7         & 93        & 94        & 6     \\
PBESol      & 97        & 3         & 76        & 24    \\
\bottomrule
    \end{tabularx}
\end{table}

Another important performance metric is dynamical stability, a crucial stability descriptor utilized by many high-throughput searches of inorganic materials~\cite{Haastrup2018,Zhu2024,Choudhary2020,Cerqueira2023}. A compound is \revision{dynamically} stable when it is in a true minimum of the potential energy surface and not in a maximum or a saddle point. In practice, it is assured by the absence of imaginary phonon frequencies in the spectrum. Unfortunately, it is well known that numerical inaccuracies often lead to small imaginary frequencies close to the $\Gamma$-point. To avoid this problem, we consider a structure to be dynamically stable if frequencies are all real across the Brillouin zone except at $\Gamma$ where we allow the three acoustic modes to have small imaginary frequencies (with a threshold of $-50$~K). This criterion was applied to all $q$-points commensurate to the supercell matrix (but not to the interpolated $q$-points).

The elements of the confusion matrix, when compared to the PBE, are listed in \cref{tab:confusionmatrix}.
Most compounds that are stable in the PBE are also stable in PBEsol and vice-versa, with the differences coming mostly from the difficulty associated to small imaginary frequencies as mentioned above. \change{MatterSim-v1} and \change{MACE-MP-0} are the most reliable with a percentage of true positives of 95\%. M3GNet, \change{SevenNet-0} and CHGNet are somewhat less accurate, especially in what concerns the percentage of true positives. Finally, the \change{eqV2-M} and ORB models perform very poorly, with more than 80\% of the unstable systems being false negatives. 

\section{Discussion}
\label{sec:conclusion}

We created a dataset that includes phonon properties of almost 10\,000 semiconductors obtained with DFT. These calculations were performed with the PBE approximation, the same approximation employed in the datasets used for the training of uMLIPs. This allows us to benchmark, without ambiguities, phonon properties calculated with uMLIPs.

In what concerns the equilibrium geometry, ORB and \change{eqV2-M} are extremely accurate and convincingly outperform all other models. This can be understood from the fact that the models output both the energy and the forces, and are trained in a very large dataset, leading to very small errors at equilibrium. Regarding phonons, however, the situation is completely different. ORB and \change{eqV2-M} yield very low quality phonons, often imaginary. We believe that the reason for this problem \revision{is that these models are non-conservative}. In fact, contrary to all other models, in ORB and \change{eqV2-M} the forces are not calculated by performing the derivative of the energy with respect to the atomic positions, but they are output directly by the network. This avoids the costly computational step of evaluating the derivatives though back-propagation and the extra freedom allows for a more accurate prediction of energy and forces. Unfortunately, it also leads to inevitable errors especially for the small displacements required for the calculation of phonons. \revision{This problematic behavior has also been reported and analyzed in~\cite{bigi2025}}. The problem can be alleviated, but far from resolved, by using larger displacements in the frozen-phonon workflow. Of course, this can lead to further problems, such as the overestimation of the anharmonic contributions.

Phonon properties calculated with \change{MatterSim-v1}, and to a lesser extent \change{SevenNet-0}, are of very high quality. Other models fare somewhere in between, exhibiting both a larger dispersion of the errors, and systematic deviations with respect to the reference PBE values.

We should note that not only the performance of the models, but also their computational efficiency, should be taken into account when choosing a uMLIP for a specific application. From the models tested here, M3GNet is by far the fastest, running in a single CPU core more efficiently than any of the other models in a full GPU. On the other extreme we have \change{eqV2-M} and \change{MACE-MP-0}, convincingly the slowest of the pack, while the rest of the models fall in between.

Our benchmark highlights the importance of considering specific optimization goals for individual metrics and understanding the trade-offs involved. Furthermore, it shows that uMLIPs are ready to be used not only for the calculation of geometries and energies, but also of response properties, that are essential for a variety of material applications. We hope our critical assessment of phonon properties will guide future training efforts and encourage the use of our dataset to further develop uMLIPs.

\section{Methods}
\label{sec:method}
\subsection{Ab initio Dataset}
To recalculate the MDR dataset~\cite{MDR_database} with PBE~\cite{Perdew1996} exchange-correlation functional, we used the code \textsc{vasp}~\cite{Kresse1996_1,Kresse1996_2}. We used all parameters consistent with the MDR dataset, with the exception of the approximation to the exchange-correlation functional that was changed from PBEsol to PBE. We followed the same workflow as MDR, but before the stringent geometry relaxation we applied a pre-relaxation step with energy and force convergence criteria of $10^{-7}$~eV/cell and $10^{-5}$~eV/\AA, respectively. For the stringent relaxation step, in accordance to the MDR calculations, we used a higher energy and force convergence criteria of $10^{-8}$~eV/cell and $10^{-8}$~eV/\AA, respectively. Next, the force constants were obtained by applying the finite displacement method as implemented in the \textsc{phonopy} python package~\cite{togo_first-principles_2023,togo_distributions_2015}.

\subsection{uMLIP evaluation}
\begin{table}[htb!]
\caption{List of models with number of training data ($N_\text{training}$), data source, and number of parameters of the models ($N_w$)}
\centering
\begin{tabular*}{\columnwidth}{@{\extracolsep{\fill}} l r l r }
\toprule
Model     & $N_\text{training}$ & Source & $N_w$ \\
\midrule
M3GNet    & 188~~k   & MPF         & 228~~k \\
\change{MACE-MP-0}      & 1.58~M  & MPtrj       & 4.69~M\\
CHGNet    & 1.58~M  & MPtrj       & 413~~k \\
\change{MatterSim-v1} & \change{6}~M    & MatterSim   & \change{4.5}~M \\
\change{SevenNet-0}  & 1.58~M  & MPtrj       & 842~~k \\
ORB       & 1.58~M  & MPtrj       & 25.2~M\\
\multirow{3}{*}{\change{eqV2-M}} & \multirow{3}{*}{110~M}     & MPtrj,       &\multirow{3}{*}{86~M} \\ 
                          &                          & Alexandria,  & \\
                          &                          & OMat         &        \\
\bottomrule
\end{tabular*}
\label{tab:pretrained}
\end{table}
For all the uMLIP models, we perform the geometry relaxation and force set calculations starting from the PBE geometry using the Atomic Simulation Environment (ASE)~\cite{HjorthLarsen2017}. To keep the space group symmetry of the PBE structure, we employ the ASE symmetrizer \textsc{FretchCellFilter}. The structure optimization is done using the fast inertial relaxation engine (FIRE)~\cite{Bitzek2006}, with force convergence criteria set to $0.005$~eV/\AA\ for all models. 

To calculate the thermodynamic properties, i.e. the vibrational entropy, the Helmhotz free energy, and the heat capacity, the phonon density of states is obtained by Fourier interpolation from the coarse calculated $q$-grid into a denser $20\times 20\times 20$ grid as in the MDR database. We set a temperature of 300~K to compute the thermodynamic properties.

\revision{The phonon density of states has been calculated using the same grid as that employed for the thermal properties. Values that are below 0.1~states/THz from PBE and model prediction were removed.
The sound velocity is calculated using group velocities near the $\Gamma$ point, that are calculated using small $q$-vectors oriented along each axis (x, y, and z). For each phonon branch, we extract the directional component of the group velocity corresponding to the axis along which the q-vector was oriented, specifically the xx, yy, and zz components. We then calculate the average of these directional components across all acoustic branches to obtain the average sound velocity.}

All models considered in this paper are open source. In ~\cref{tab:pretrained} we list their training set sizes, data sources, and the number of parameters.

\section{Data Availability}
\label{sec:data_ava}
The phonon dataset is available in \textsc{alexandria} which can be accessed and/or downloaded from \url{https://alexandria.icams.rub.de/} under the terms of the Creative Commons Attribution 4.0 License.  

\section{Code Availability}
\label{sec:code_ava}
All code developed in this work is freely available at  \url{https://github.com/hyllios/utils/tree/main/}.

\section{Acknowledgements}
A.L. and M.A.L.M. acknowledge funding from the Horizon Europe MSCA Doctoral network grant n.101073486, EUSpecLab, funded by the European Union. S.B. and D.S. acknowledge financial support from the Deutsche Forschungsgemeinschaft (DFG, German Research Foundation) through the project BO4280/11-1. H.C.W and M.A.L.M would like to thank the NHR Centre PC2 for providing computing time on the Noctua 2 supercomputers.

\section{Author Contributions}
A.L. and M.A.L.M developed the high-throughput workflow;  A.L. , D.S.,  and  M.A.L.M performed the response calculations; A.L. and M.A.L.M. performed the machine learning validations;  H.-C. W., S.B., and M.A.L.M directed the research.  All authors participated equally in the interpretation of the results and in the writing of the manuscript.

\section{Competing Interests}
S.B. declares to be an Associate Editor for npj Computational Materials. This role has not influenced the peer review or editorial process for this manuscript, which has been handled independently according to the journal's standard procedures for manuscripts with editorial board member authorship.

\bibliography{bib.bib}

\end{document}